# Detection of Node Clones in Wireless Sensor Network Using Detection Protocols

Neenu George[#1], T.K.Parani[#2]

[#1]II Year M.E Student, [#2]Assistant Professor

[#1, #2] ECE Department, Dhanalakshmi Srinivasan College of Engineering,

Coimbatore,tamilnadu, India.

*Anna University*

Email: neenugeorge001@gmail.com and parani30@gmail.com

**Wireless sensor networks consist of hundreds to thousands of sensor nodes and are widely used in civilian and security applications. One of the serious physical attacks faced by the wireless sensor network is node clone attack. Thus two node clone detection protocols are introduced via distributed hash table and randomly directed exploration to detect node clones. The former is based on a hash table value which is already distributed and provides key based facilities like checking and caching to detect node clones. The later one is using probabilistic directed forwarding technique and border determination. The simulation results for storage consumption, communication cost and detection probability is done using NS2 and obtained randomly directed exploration is the best one having low communication cost and storage consumption and has good detection probability.**

## I. INTRODUCTION

A wireless sensor network (wsn) is a high and new technology consists of spatially distributed autonomous sensors to monitor physical or environmental conditions and to pass data through the network to a main location. It is built of hundreds or thousands of nodes and each node act as sensor. Wireless sensor network consists of base stations and number of wireless sensors. These sensors node network has transceiver, micro controller, electronic circuit and energy source. Sensor networks have significant constrains and the individual sensor nodes are typically inexpensive, tiny, distributed, low power and low complexity nodes which used lightweight processors and cheap hardware components of low tamper résistance. And these sensor nodes are often deployed in hostile environments, are highly independent and require only a minimum amount of supervision. The cost of these sensor nodes depend on resources such as energy, memory, speed, bandwidth etc..They are widely used in physical and environmental situations. The wireless sensor network avoids the use of lot of wiring, can accommodate any devices at any time, and is accessed through centralized monitor and highly flexible [8]. The main goal of wsn is to reduce power consumption and to optimize computing resources.

The bandwidth range of wsn is radio frequency. Wsn are ad hoc networks (wireless nodes that self organize into an infrastructure less network). In contrast to other adhoc network, wsn need essentially sensing and data processing. It has many more nodes and is densely deployed [8]. Hardware must be cheap and nodes are more prone to failures. Communication scheme is many to one (data collected at base station) rather than peer to peer and nodes are static.

The main problems deals with wsn are easy to hack, low speed of communication, high cost and interference. Because of this hackers many attacks affects the wsn. Among those attacks serious and dangerous one is *node clone attack*. In this attack the adversary may capture some nodes in the network when they are in hostile environment and extract the secret credentials data and information from nodes, reprograms or modifies the data and creates replicas or clones of such nodes in the network. Then these compromised nodes plays active in network and thus the adversary may gain the control over the network [1]. Thus security of network had lost and more over these cloned nodes can create more attacks like DoS inside the network which corrupts the information [2]. If these clones are left undetected, the network is unshielded to attackers and thus extremely vulnerable [3].

Therefore in this paper, an effective two novel node clone detection protocols are proposed to detect the node clones. The previous works incurs more communication cost and required to transmit more messages resulted in the reduction of life time. The first one is *distributed hash table* (DHT) which is based on hash table value of (key, record) by which a fully decentralized, key-based caching and checking system is constructed to catch nodes. DHT enables sensor nodes to construct the over lay network. The key plays vital role in DHT mechanism which determines the destination node of the message [1]. But this DHT incurs same communication cost as previous, have some storage consumption and strong detection probability. Second one is distributed detection protocol, named *randomly directed exploration (RDE)* in which probabilistic directed forwarding technique along with random initial direction and border determination. Every node contains signed version of neighbor list and the detection round is initiated by sending claiming message by the nodes to randomly selected neighbors [5]. The communication cost is reduced by using border determination. And this protocol has to store only the list of neighbor nodes so consumes less memory. So the RDE stands with low communication cost, less storage consumption and high detection probability. The

RDE and DHT protocols are only used to detect the node clones in the wsn..

The rest of paper is organized as follows. First the previous counter measures are discussed in Section II. Then, present preliminaries in Section III. After wards detailed description about DHT and its performance in Section IV. The RDE is detailed in Section V with its performance. Finally conclude the work in Section VI.

## II. PREVIOUS WORKS

The earliest method to detect node clones was *prevention schemes* and key plays the main role which provided to nodes by mobile trusted agents. The private key of node comprises of location and identity. But the problems arise here are attackers may takes some time to compromise the nodes (compromising time) in the network. As the compromising time decreases the number of clone nodes increases thus badly affects the security of the network. And also prevention scheme is applicable to only some specific applications. The assumption made on trusted agents is not too strong [7].

In the *centralized detection* method a base station is connected to each node. Each node sends a list of its neighbor nodes and location to base station. The communication cost is limited by constructing subsets of nodes. Even though communication cost is reduced the life time expectancy of the network is decreased due to the communication burden of the nodes near to the base station [4].

## III. PRELIMINARIES

### A. Network Model

The network used is of large scale and have 'n' number of resource constrained sensor nodes. Each node has unique ID and a corresponding private key. The public key $k_\alpha$ is the node ID and private key is $k_\alpha^{-1}$. Message M signed by the node α using private key is $[M] k_\alpha^{-1}$. The location and current relative time of every node is determined by secure localization protocol and secure time synchronization scheme respectively [1]. And these are not specifying since they are not so important to proposed protocols. As per previous approaches the base station is not powerful in this model, instead of that an initiator plays as a trusted role for initiating the detection round procedures. During node clone detection the sensor nodes are assumed to be stationary. So the node clone can be determined by the collision of location for one node ID [5].

### B. Adversary Model

The sensor networks are more vulnerable to attacks in hostile environment. The adversary can capture some nodes, can modify or reprogram it and obtains all the secret credentials data. Thus the compromised node creates replicas or clones of such mischievous nodes and adversary may gain control of the whole network by deploying these replicas in place that are decided intelligently. Adversary is always aware of detection protocol and manages to conceal the existence of clone [5]. Adversary interferes with the detection scheme in three ways.

First, cloned nodes may not participate in the detection rounds. Second, cloned nodes may drop or modify the messages. Lastly they take some time to compromise the nodes is limited [1].

### C. Performance Metrics

The performance metrics used to compare both protocols are
(i) *Communication cost*: the average number of messages sent per node is used to represent Communication cost.
(ii) *Storage consumption:* low cost sensors have limited amount memory. Average cache table size per node represents storage consumption.
(iii) *Detection probability:* average number of witness nodes per node represents detection probability.

## IV. DISTRIBUTED HASH TABLE

Distributed Hash Table is the node clone detection protocol which provides decentralization scheme with the key based caching and checking. Distributed Hash Table is based on a hash table of (key, record) pair which is already distributed. The distributed hash table enables the sensor nodes to form an overlay network. The key plays vital role in distributed hash table and key determines where to send the message from source node i.e. the destination node is determined by the key and source doesn't know anything about the destination node. The detection round initiated by initiator by sending an action message (involves nonce, seed, and time). Then every observer nodes constructs claiming message for each neighbor node, referred as examinee and sends the message with probability $p_c$ to reduce the communication over work. The key which determine the destination node of message is the hash value of concatenation of seed and examinee ID [1]. During distributed hash table detection round a claiming message will transmitted to destination node which will cache ID- location pair and check for node clone detection.

Distributed Hash Table is a decentralized distributed system which provides a key based look up service. (Key, record) pairs are stored in the table any active node can store and retrieve records associated with specific keys. Thus distributed hash table maintain mapping from keys to records among nodes. Chord is used and choose chord as a distributed hash table implementation to demonstrate protocol. Massive virtual ring is formed by chord in which every node is located at one point, and owning a segment of the periphery. Hash function is used to achieve pseudo randomness on output by mapping an arbitrary input into a b-bit space (in the ring).Chord coordinate is assigned for each node and can join the network. Here a node's Chord point's coordinate is the hash value of the node's MAC address [1].one segment that ends at the node's Chord point is related to every node, and all records whose keys fall into that segment will be transmitted to and stored in that node[5].Every node maintains a finger table of size t= O (log n) to facilitate a binary-tree search. The finger table for a node with responsible for holding the t keys.

TABLE I
DISTRIBUTED DETECTION PROTOCOLS COMPARISON, WHERE n IS NETWORK SIZE, d NODE DEGREE

| Protocols | Nodes requirements | Communication cost | Memory cost | Detection Cost |
|---|---|---|---|---|
| Node to network broadcasting | Neighbors information | $O(n)$ | $O(d)$ | Strong |
| Randomized multicast | All nodes data | $O(n)$ | $O(d\sqrt{n})$ | Acceptable |
| Line selected | All nodes data | $O(\sqrt{n})$ | $O(d\sqrt{n})$ | Acceptable |
| RED | Knowledge of network geography | $O(\sqrt{n})$ | $O(d\sqrt{n})$ | Strong |
| DHT | DHT nodes information | $O(\log n \sqrt{n})$ | $O(d)$ | Strong |
| RDE | Neighbors information | $O(\sqrt{n})$ | $O(d)$ | Good |

between 10 and 20. The DHT enable sensor nodes to construct a chord overlay network. Cloned node may not participate in this overlay network construction[1]. And this overlay network construction is independent of node clone detection. Nodes possess the information of their direct predecessor and successor in the Chord ring and also caches information of its consecutive successors in its *successors table[6]*. The communication cost is thus reduced by this cache mechanism and it enhances systems robustness. Selection of inspectors is done using the facility of the successors table.

*Detection round stages*

(i) The initial stage of detection round is done by activating all nodes by releasing an action message by initiator

$M_{ACT}$=nonce, seed, time, {nonce||seed||time} $k^{-1}_{initiator}$

During each rounds the value of nonce increases monotonously and it intended to prevent the DoS attacks.

(ii) By receiving the action message each node verifies the value of nonce with previous values and verifies the signature of the message. If both are valid node will updates the nonce and stores the seed. The node act as observer to generate claiming message for each neighbor at the designated action time and transmits the message through the overlay network with respect to the claiming probability $p_c$.

$M_{\alpha 4\beta}$=$id_\beta$, $L_\beta$, $id_\alpha$, $L_\alpha$, { $id_\beta$ ||$L_\beta$ ||$id_\alpha$ ||$L_\alpha$||nonce} $k^{-1}_\alpha$.

where, $L_\alpha$, $L_\beta$ are locations of α and β, respectively.

(iii) Chord intermediate nodes will forwards claiming message to its destination node. Only the source node, Chord intermediate nodes, and the destination node need to process a message, whereas other nodes along the path simply route the message to temporary targets. Algorithm 1 for handling a message and If the algorithm returns NIL, then the message has arrived at its destination. Else the message will forwarded to the next node with the ID that is returned by Algorithm[1].

**Algorithm 1:**

dht_ handle message($M_{\alpha 4\beta}$) handle a message in the DHT-based detection, where y is the current node's Chord coordinate, finger[i] is the first node on the ring that succeeds key((y+$2^{b-1}$ mod $2^b$),I £ [1,t] ,successors [j] is the next $j^{th}$ successor j £[1,g][1].

**Output:** NIL if the message arrives at its destination; otherwise, it is the ID of the next node that receives the message in the Chord overlay network[1].

1: key<=H (seed||$id_\beta$)
2: **if** key £ [predecessor] **then** {has reached destination}
3: inspect $M_{\alpha 4\beta}$ {act as an inspector, see Algorithm 2}
4: **return** NIL
5: for i=1 to g **do**
6: **if** key £(y, successors [i]) **then** {destination is in the next Chord hop}
7: inspect $M_{\alpha 4\beta}$ {act as an inspector, see Algorithm 2}
8: **return** successors [i]
9: **for j= 1** to t **do** {**for** normal DHT routing process}
10: **if key** £ [(y+$2^{b-1}$ mod $2^b$,y)], **then**
11: **return** finger [j]
12: **return** successor [g]

**Algorithm 2:** inspect $M_{\alpha 4\beta}$: Inspect a message to check for clone detection in the DHT-based detection protocol
1: verify the signature of $M_{\alpha 4\beta}$
2: **if** $id_\beta$ found in cache table **then**
3: **if** $id_\beta$ has two distinct locations {found clone, become a witness}
4: broadcast the evidence
5: **else**

6: buffer $M_{\alpha 4\beta}$ into cache table

Message for node clone detection is examined by Algorithm 1 and Algorithm 2 compares the message with previous inspected messages that are buffered in the cache table[1]. All records in the cache table should have different examinee ID. If there exist two messages $M_{\alpha 4\beta}$ and $M_{\alpha' 4\beta'}$ satisfying $id_\beta = id_{\beta'}$ and $L_\beta \neq L_{\beta'}$ shows that exists clone and then the witness node broadcasts the evidence to notify the whole network. All integrity nodes verify the evidence message and stop communicating with the cloned nodes. The witness does not need to sign the evidence message.

### D. Performance Analysis of DHT

*Communication cost*: The average path length between two random nodes by l which varies from $O(\log n)$ to $O(\sqrt{n})$. On the basis of Chord's properties the number of transfers in the Chord overlay network is $c \log n$, where c is a constant number, usually less than 1. Therefore, the average path hop length of a message is $cl \log n$[1]. There are $p_c dn$ claiming messages in total for a round of detection. Thus shown in fig 1(a) the average number of messages sent per node is given by $p_c dcl \log n$. Since the $p_c$, d, and c are constant, the asymptotic communication cost of the DHT-based protocol is between $O(\log^2 n)$ and $O(\sqrt{n}\log n)$.

*Storage Consumption*: In particular, protocol shows strong resilience against message-discarding by cloned nodes. In fact, the more cloned nodes, the less the size of cache tables for integrity nodes as storage consumption and the more witnesses as security level shown in figure 2 (a). Good pseudo-randomness of the Chord system, on average, every node stores one record in its cache table associated with one examinee's ID as its destination, regardless of the number of claiming messages per examinee. Let $p_r$ denote the probability of a predecessor receiving a specific claiming message, then the probability of a predecessor holding a record for an examinee is $1-(1-p_r)^m$. Average cache table size $s=1+g(1-(1-p_r)^m)$[1].

*Detection probability*: Even if there are 10% nodes that maliciously discard messages, the number of witnesses is pretty high. The g predecessor nodes of the destination may become witnesses if and only if they receive at least two claiming messages associated with different cloned nodes. Average witness number $w=1+g(1-(1-p_r)^m)^2$. In an ideal case Average cache table size $s =1+gm/g+m$ where there are m independent claiming messages for each examinee and g is the successors table size. The average witness number, when there are two cloned nodes, is $w=1+2gm^2/(g+m)(g+2m)$[1].

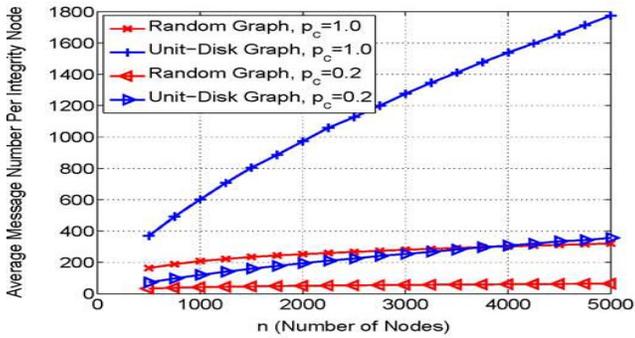

(a)

Figure 1. simulation results of DHT detection on number of nodes

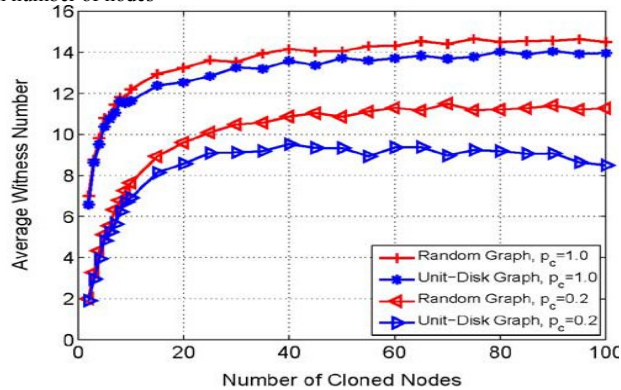

(a)

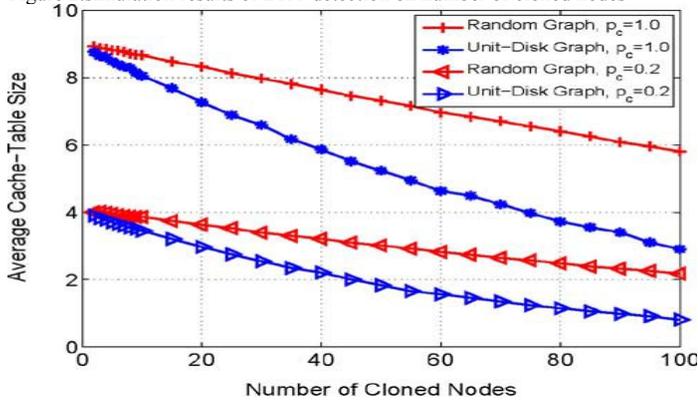

Figure 2.simulation results of DHT detection on number of cloned nodes

(a)

Figure 3.simulation results of DHT detection on number of cloned nodes

## V. RANDOMLY DIRECTED EXPLORATION

The problems associated with the dht are it incurs more communication cost because of the chord overlay network and thus it is sensitive to energy and storage consumption. To overcome these problems a new node clone detection protocol introduced namely randomly directed exploration. Here the each node only needs to know and buffer a neighbor list having all neighbors ID and locations. During detection round each node constructs claiming message with signed version of neighbor list and then deliver message to others which will compares with its own neighbor list to detect node clone. If there exists any node clone, one witness node successfully catches the clone and notifies the entire network by broadcasting. The efficient way to achieve randomly directed exploration needs some mechanisms and routing protocols. First the claiming message needs to provide maximum hop limit and it is sent to random neighbors. Then the further message transmission will maintain a line and this transmission line property enables a message to go through a network as fast as possible[6]. The communication cost of this protocol is low and it is limited by the border determination mechanism. And the assumption made here is that each node knows about its neighbors locations.

*Detection round*

Initially the node clone detection round is activated by the initiator. At the right mentioned action time, each node creates its own neighbor list (ID of neighbor and location). Then that node act as an observer for all its neighbors and starts to generate claiming messages. The claiming message involves node ID, location and its neighbor list[6]. The claiming message by node is constructed by

$M_\alpha$=ttl, id$\alpha$, L$\alpha$, neighbor list where ttl is time to live.

**Algorithm 3:** rde-processmessage M$\alpha$: An intermediate node processes a message
1: verify the signature of M$\alpha$
2: compare its own neighbor-list with the neighbor-list in M$\alpha$
3: **if** found clone **then**
4: broadcast the evidence;
5: ttl<=ttl-1
6: **if** ttl $\leq$ **0 then**
7: discard M$\alpha$
8: **else**
9: next node<=get next node (M$\alpha$) {See Algorithm 4}
10: **if** next node =NIL **then**
11: discard M$\alpha$
12: **else**
13: forward M$\alpha$ to next node[6]

The intermediated nodes will change the value of ttl during transmission. In each time, the node transmits message to a random neighbor. When an intermediate node $\beta$ receives a claiming message $M_\alpha$, it launches rde-processmessage M$\alpha$. During the processing the node clone is detected by comparing the neighbor list of node which acts as inspector $\beta$ with neighbor list in the message. If clone detected then the witness node $\beta$ will broadcast an evidence message M $_{evidence}$= (M$_\alpha$,M$_\beta$) to notify the whole network such that the cloned nodes are removed from the network[6]. Node decreases the message's ttl by 1 and discards the message if ttl reaches zero during routing; otherwise it will query Algorithm 4 to determine the next node receiving the message.

**Algorithm 4:** get next node (Mα): To determine the next node that receives the message
1: determine ideal angle, target zone, and priority zone
2: **if** no neighbors within the target zone **then**
3: **return** NIL
4: **if** no neighbors within the priority zone **then**
5: next node<= the node closest to ideal angle
6: **else**
7: next node<= a probabilistic node in the priority zone, with respect to its probability proportional to angle distance from priority zone border
8: **return** next node[6].

*Deterministic directed transmission*: The ideal direction can be calculated when node receives a claiming message from previous node and the next destination node should be closest to the ideal direction for the best effect of line transmission. *Network border determination:* The communication cost is reduced by taking network shape into consideration. Due to physical constrains in many sensor network applications, there exist outside borders. The claiming message can be directly discarded when reaching some border in the network. To determine a target zone then no neighbor is found in this zone, target range is used along with ideal direction, the current node will conclude that the message has reached a border, and thus throw it away. *Probabilistic directed transmission*: priority range along with the ideal direction is used to specify a priority zone, in which the next node will be selected. The deterministic directed candidate within the target zone will be selected as the next node when no nodes are located in that zone,. If there are several nodes in the priority zone, their selection probabilities are proportional to their angle distances to priority zone border. As a result, to reduce detection probability dramatically the adversary may remove some nodes in strategic locations Claiming messages transmissions from a cloned node's neighbors are highly correlated, which affects the protocol communication and security performance[1]. Those drawbacks are overcome, by the probabilistic directed mechanism, and the protocol performance is improved significantly

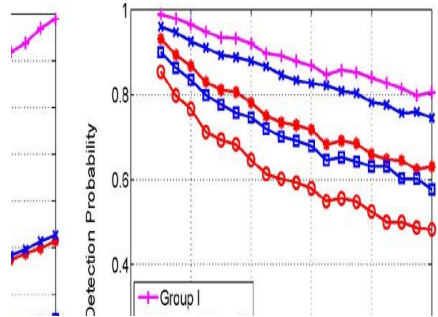

Figure 4 simulation results of RDE on varying size networ

..

*E. Performance Analysis of Communication cost*: The parameter settings. On average, and each message transmits at dense network, but ttl is because there are nodes in the it is very likely for any two network topology[6]. In other across the network. The upper exploration protocol is O$\sqrt{n}$ and its shown in fig 4 (a)

*RDE*
RDE's communication cost depends on the routing there are r claiming messages sent by each observer, most ttl hops, r is a constant small number, say 1 for a generally related to the network size. So ttl=$\sqrt{n}$ network, and by the line property of protocol routing, nodes to be reachable within $\sqrt{n}$ hops for a normal words, ttl=$\sqrt{n}$ would be sufficient for messages to go bound of communication cost in the randomly directed

*Detection probability:* Relieving message-discarding and protecting witness are achieved by random initial direction and probabilistic directed transmission. By them, there is no critical location to affect message transmission, which limits the capacity of message-discarding, and every neighbor of a cloned node has similar potential to become witness so it is hard for the adversary to get rid of witness in advance[1]. The RDE protocol's detection probability is determined by the number of nodes that are reached when randomly drawing lines where each has a random initial angular and fixed number of nodes along this direction with the border limitation. Let h denote the reachable node number; ₀, it is a function of (an initial angular), ttl (the number of maximum hops), and v (the number of the claiming messages). Therefore, for a network with n nodes, the detection probability is given by $P_{RDE}$=h (ttl,₀,v)/n shown in fig 4(b).

*Storage consumption:* The RDE protocol is exceedingly memory-efficient. It does not rely on broadcasting; thus, no additional memory is required to suppress broadcasting flood. The protocol does not demand intermediate nodes to buffer claiming messages, all memory requirement lies on the neighbor-list, which, in fact, is a necessary component for all distributed detection approaches. Therefore, the protocol consumes almost minimum memory shown in fig 4 (c).

## VI CONCLUSION

Sensor nodes lack tamper-resistant hardware and are subject to the node clone attack. So two distributed detection protocols are presented: One is based on a distributed hash table, which forms a Chord overlay network and provides the key-based routing, caching, and checking facilities for clone detection, and the other uses probabilistic directed technique to achieve efficient communication overhead for satisfactory detection probability. While the DHT-based protocol provides high security level for all kinds of sensor networks by one deterministic witness and additional memory-efficient, probabilistic witnesses, the randomly directed exploration presents outstanding communication performance and minimal storage consumption for dense sensor networks. From the analysis and simulation results, the randomly directed exploration protocol outperforms all other distributed detection protocols in terms of communication cost and storage requirements, while its detection probability is satisfactory, higher than that of line-selected multicast scheme.

## REFERENCES

[1] Zhijun Li, Member, IEEE, and Guang Gong,,".On the Node Clone Detection in Wireless Sensor Networks", in proc 5th IEEE transactions,Volume 40,no.11,pp 17-23,2013.

[2] H.Wen,J.Luo,L.Zhou,"Light weight and effective detection scheme for node clone attacks in wsn*", in proc* IET wsn,2011.

[3] Kai Xing, X.Chen,D.H.C Du,F.Liu,"Real time detection of clone attcks in wireless sensor networks", *IEEE infocom*,2008.

[4] B. Parno, A. Perrig, and V. Gligor, "Distributed detection of node replication attacks in sensor networks," in *Proc. IEEE Symp. Security Privacy*, 2005, pp. 49–63

[5] I.Stoica,R.Morris,,D.L.Nowell,D,R.Karger,M.F.Kaashoek,Hari Balakrishnan,"Chord:a scalable peer to peer look up protocol for internet applications,'*IEEE/ACM Trans.Netw.*,vol.11,no.1,pp,17-32,Feb.2003.

[6] Zhijun Li And Guang Gong, "Randomly directed exploration: an efficient node clone detection protocol in wireless sensor network" ,in proc 5 th IEEE Trans.Volume 11,pp34-4..2009.

[7] C.Bekara And M.L.Maknavicius,"A new protocol for securing wsn against node replication attacks,"in third *IEEE International Conference* on wireless and mobile computing,networking and communications,2007,pp 59-59

[8] www.wikipedia.com